\documentclass[final,3p,times,authoryear]{elsarticle}

\usepackage[T1]{fontenc}
\usepackage[utf8]{inputenc}
\usepackage{amsmath,amssymb,mathrsfs}
\usepackage{booktabs}
\usepackage{tabularx}
\usepackage{array}
\usepackage{graphicx}
\usepackage{subcaption}
\usepackage{microtype}
\usepackage{xurl}
\usepackage[hidelinks]{hyperref}

\biboptions{round,semicolon,authoryear}

\newcolumntype{Y}{>{\raggedright\arraybackslash}X}

\begin{document}

\begin{frontmatter}

\title{Herding, Momentum, and Reversal in China's A-Share Market:
An Agent-Based Network Model with Information Diffusion}

\author[ruc]{Jiahao Weng\corref{cor1}}
\cortext[cor1]{Corresponding author}
\ead{author.email@institution.edu}

\affiliation[ruc]{
  organization={Department of Physics, Renmin University of China},
  city={Beijing},
  postcode={100872},
  country={China}
}

\begin{abstract}
This study develops an agent-based financial market model to explain stock-price
momentum and reversal through the joint effects of local herding and delayed
information diffusion. Investors form heterogeneous Gaussian beliefs about the
next-period price, choose among buying, selling, and remaining inactive, and
revise their action probabilities in response to neighboring investors. The
local interaction structure is represented by von Neumann and Moore lattices
and is later replaced by Erd\H{o}s--R\'enyi and Watts--Strogatz networks for
robustness. A separate information process updates investor beliefs through a
finite-speed diffusion mechanism, allowing informational adjustment to be
distinguished from behavioral imitation. The simulations show that stronger
herding produces spatially clustered trading, larger price fluctuations, and
more pronounced excess kurtosis in returns. Faster information diffusion
reduces the time required for prices to approach the signal-implied value,
whereas the combination of information diffusion and social reinforcement
generates overshooting and subsequent reversal. An empirical application to
China's A-share market compares conventional CSAD and LSV measures with a
rolling tail-based herding indicator obtained after Johnson \(S_U\)
transformation. The indicators display similar time variation and rise during
major market disruptions. These findings identify information delay, local
social reinforcement, and the eventual decay of herding as complementary
mechanisms behind momentum and reversal.
\end{abstract}


\end{frontmatter}

\section{Introduction}
\label{sec:introduction}

Financial markets aggregate heterogeneous expectations into prices. In the
benchmark efficient-market view, prices incorporate available information and
past returns should not systematically predict future excess returns
\citep{fama1970}. In practice, however, limits to arbitrage and correlated
non-fundamental demand can allow prices to depart from fundamental values
\citep{delong1990}. Momentum and reversal are two prominent manifestations of
this departure: recent winners and losers may continue along the same path over
an intermediate horizon, but sufficiently large deviations are often followed
by corrections.

Herding provides a behavioral channel through which individually modest
decisions become an aggregate price force. Investors may imitate observed
actions because private information is incomplete, because they infer that
other traders possess superior information, or because reputational and social
pressures make conformity attractive \citep{banerjee1992,bikhchandani1998,
scharfstein1990}. Once imitation becomes self-reinforcing, an information
cascade can suppress private signals and reduce the informational content of
prices \citep{hirshleifer2003}. Herding has been documented across asset
classes and institutional settings, including equity, mutual-fund, shipping,
and cryptocurrency markets \citep{wermers1999,papapostolou2017,
messis2023,sharma2024}.

The mechanism is especially relevant to China's A-share market, where retail
participation and uneven information-processing capacity make local social
interaction potentially important. Existing empirical measures capture
different aspects of the phenomenon. Cross-sectional dispersion measures
detect whether individual returns converge toward the market return during
large market moves \citep{christie1995,chang2000}. The LSV measure instead
identifies unusually concentrated buy or sell decisions by institutional
investors \citep{lakonishok1992}. These approaches are useful, but they are
usually estimated over a window of observations and do not by themselves
separate common reactions to information from imitation.

This paper addresses that distinction with an agent-based market in which
belief updating and social reinforcement are modeled as separate processes.
Investors first translate subjective price distributions into buy, sell, and
hold probabilities. Their probabilities are then shifted toward the recent
actions of network neighbors. Independently, a signal diffuses through the
network and changes the beliefs of investors it reaches. Market clearing maps
the resulting heterogeneous decisions into a price path.

The paper makes three contributions. First, it embeds local interaction in a
market-clearing model rather than imposing a representative herding
coefficient at the aggregate level. Second, it separates delayed information
diffusion from imitation, thereby clarifying why both rational underreaction
and behavioral reinforcement can produce momentum. Third, it proposes a
rolling distributional indicator based on transformed return tails and compares
it with established CSAD and LSV measures in the Chinese market. The design
does not claim that every momentum or reversal episode is caused by herding;
instead, it provides a disciplined mechanism through which herding can amplify
informational price adjustment.

The remainder of the paper is organized as follows. Section
\ref{sec:literature} reviews the relevant literature. Section
\ref{sec:model} presents the agent-based model and market-clearing condition.
Section \ref{sec:information} introduces information diffusion. Section
\ref{sec:simulation} reports the simulation design and mechanisms. Section
\ref{sec:empirical} describes the empirical measures and application. Section
\ref{sec:robustness} examines alternative network structures, and Section
\ref{sec:conclusion} concludes.

\section{Related literature}
\label{sec:literature}

\subsection{Herding and information cascades}

Sequential-decision models show that rational agents can ignore their private
signals when the observed actions of predecessors are sufficiently informative
\citep{banerjee1992,bikhchandani1998}. In financial markets, reputational
concerns can produce similar behavior among professional investors
\citep{scharfstein1990}. Herding may therefore be intentional, when traders
explicitly infer information from others, or unintentional, when common
constraints and public signals cause correlated decisions
\citep{papapostolou2017}. This distinction matters empirically because common
responses to fundamentals do not necessarily imply irrational imitation.

Social and professional connections provide a natural channel for information
and behavior to spread. Portfolio-manager connections predict investment
decisions and returns \citep{cohen2008}, while transaction-level evidence shows
that central investors in empirical trading networks act earlier around
information events \citep{ozsoylev2014}. These findings motivate the use of a
local network rather than an all-to-all interaction rule.

\subsection{Momentum, overreaction, and reversal}

Delayed information diffusion can generate return continuation even when each
investor reacts rationally to the information received. In
\citet{hong1999}, news reaches informed agents gradually, while momentum
traders extrapolate observed price changes. The interaction creates initial
underreaction, later continuation, and possible overreaction. Herding adds a
second source of persistence: recent actions alter the decisions of connected
investors and reinforce an existing trend. Once the local majority saturates,
the arrival of offsetting information or the weakening of imitation can
produce reversal.

Institutional trading offers related evidence. Mutual funds display correlated
trading and momentum-oriented behavior \citep{grinblatt1995,wermers1999}.
These results do not establish that all correlated trading is irrational, but
they indicate that position changes can transmit and amplify market trends.

\subsection{Empirical measurement of herding}

\citet{christie1995} propose cross-sectional return dispersion as an aggregate
herding diagnostic. \citet{chang2000} replace standard deviation with the
cross-sectional absolute deviation (CSAD) and test for a nonlinear relation
between dispersion and market returns. A significantly negative coefficient on
the squared market return is interpreted as evidence that individual returns
converge during large market moves. International evidence suggests that the
strength and sign of this relation vary across markets and regimes
\citep{chiang2010}.

The LSV measure takes a different perspective and asks whether the fraction of
buyers in a stock differs from the market-wide buying probability by more than
expected under independent trading \citep{lakonishok1992}. Its direct use of
trades is attractive, but the adjustment requires sufficiently many
transactions and cannot fully distinguish imitation from correlated
information. The distributional indicator introduced in Section
\ref{sec:tailindicator} is intended as a high-frequency complement, not a
replacement for CSAD or LSV.

\section{Agent-based market model}
\label{sec:model}

\subsection{Investor beliefs and actions}

The market contains \(N\) investor agents connected by a graph
\(\mathcal{G}=(\mathcal{V},\mathcal{E})\). An agent may represent an individual
investor or a cluster with internally similar beliefs. Let \(X_t>0\) denote the
asset price at date \(t\). Conditional on \(X_t\), investor \(i\)'s prior belief
about the next-period price is Gaussian:
\begin{equation}
f_{i,t+1}(x\mid X_t)
=
\frac{1}{\sqrt{2\pi}\,\widehat{\sigma}_{i}}
\exp\left[
-\frac{(x-\widehat{x}_{i})^2}
{2\widehat{\sigma}_{i}^{\,2}}
\right],
\label{eq:prior}
\end{equation}
where \(\widehat{x}_{i}\) and \(\widehat{\sigma}_{i}\) are the subjective mean
and standard deviation.

The action set is
\(\mathcal{A}=\{B,S,H\}\), denoting buy, sell, and hold. For a candidate
next-period price \(x\), the baseline sell and buy propensities are
\begin{align}
\pi_{i,t}^{S}(x)
&=
\frac{1}{1+\exp\{-k[x-(1+\Delta)\widehat{x}_{i}]\}},
\label{eq:sellprob}\\
\pi_{i,t}^{B}(x)
&=
\frac{1}{1+\exp\{k[x-(1-\Delta)\widehat{x}_{i}]\}},
\label{eq:buyprob}
\end{align}
and
\begin{equation}
\pi_{i,t}^{H}(x)
=1-\pi_{i,t}^{B}(x)-\pi_{i,t}^{S}(x).
\label{eq:holdprob}
\end{equation}
Here \(k>0\) governs sensitivity to perceived mispricing and \(\Delta>0\)
defines an inaction band around the subjective mean. Parameters must be chosen
so that Eq.~\eqref{eq:holdprob} remains nonnegative over the relevant price
range.

\begin{figure}[hbt]
    \centering
    \includegraphics[width=0.5\linewidth]{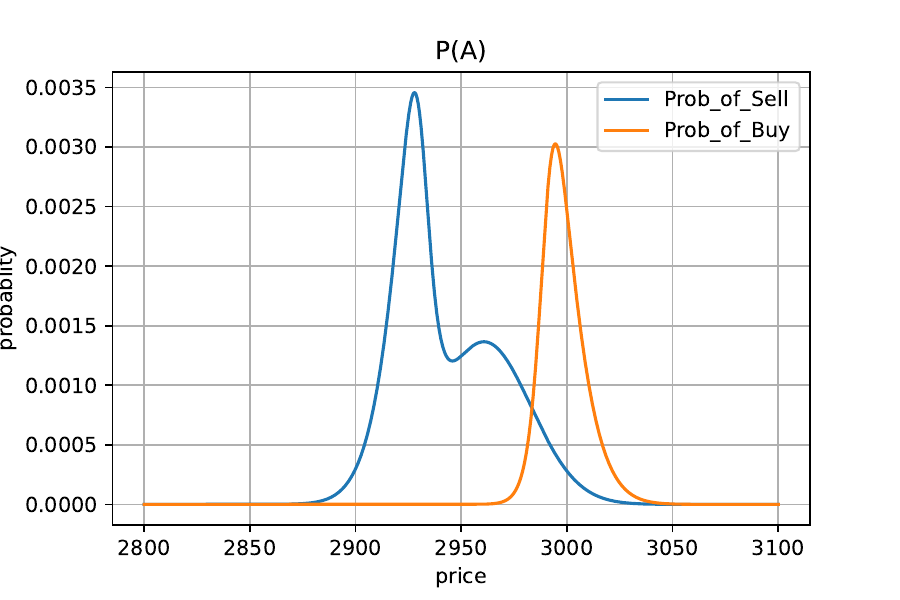}
    \caption{Buy and sell probabilities conditional on the candidate price.}
    \label{fig:PA}
\end{figure}

\subsection{Local herding}

Let \(\mathcal{N}_i\) denote investor \(i\)'s neighbors and
\(d_i=|\mathcal{N}_i|\). If \(n_{i,t}^{a}\) neighbors chose action
\(a\in\mathcal{A}\) at date \(t\), their observed action share is
\begin{equation}
h_{i,t}^{a}=\frac{n_{i,t}^{a}}{d_i}.
\label{eq:neighborshare}
\end{equation}
The decision probability after social interaction is a convex combination of
the agent's own propensity and the neighbor share:
\begin{equation}
\widetilde{\pi}_{i,t}^{a}(x)
=(1-\gamma)\pi_{i,t}^{a}(x)+\gamma h_{i,t}^{a},
\qquad
a\in\mathcal{A},
\label{eq:herding}
\end{equation}
where \(\gamma\in[0,1]\) is the herding intensity. This specification preserves
normalization and gives \(\gamma\) a direct interpretation: it is the weight
placed on locally observed behavior.

The baseline simulations use periodic lattices to remove boundary effects. A
von Neumann lattice connects each node to its four orthogonal neighbors,
whereas a Moore lattice adds the four diagonal neighbors. Both networks are
regular, but the Moore network transmits actions and information more quickly
because each investor has twice as many local contacts.

\begin{figure}[hbt]
    \centering
    \begin{subfigure}{0.3\textwidth}
        \includegraphics[width=\textwidth]{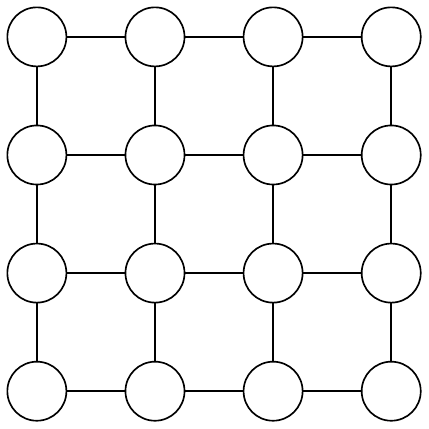}
        \caption{Von Neumann network}
        \label{fig:von-neumann}
    \end{subfigure}
    \begin{subfigure}{0.3\textwidth}
        \includegraphics[width=\textwidth]{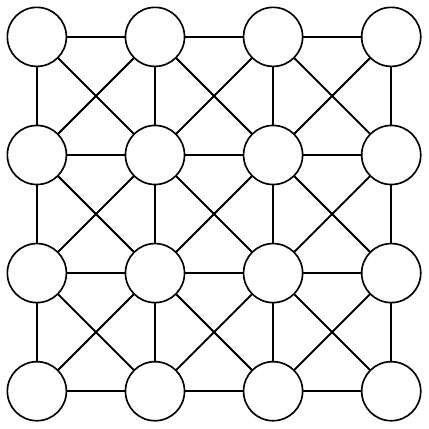}
        \caption{Moore network}
        \label{fig:moore}
    \end{subfigure}
    \caption{\(4\times4\) von Neumann and Moore networks.}
    \label{fig:lattices}
\end{figure}

\subsection{Positions and market clearing}

Investor \(i\) chooses a risky position \(q_i\) to balance expected excess
payoff and variance:
\begin{equation}
\max_{q_i}\;
\mathcal{U}_{i,t}
=q_i\left[
\mathbb{E}_{i,t}(X_{t+1})-X_t e^{r_f}
\right]
-\lambda_i q_i^2
\mathbb{V}_{i,t}(X_{t+1}),
\label{eq:utility}
\end{equation}
where \(r_f\) is the one-period risk-free rate and \(\lambda_i>0\) is risk
aversion. The unconstrained optimal position is
\begin{equation}
q_{i,t}^{*}
=
\frac{\mathbb{E}_{i,t}(X_{t+1})-X_t e^{r_f}}
{2\lambda_i\mathbb{V}_{i,t}(X_{t+1})}.
\label{eq:position}
\end{equation}

For a candidate price \(x\), expected signed demand is
\begin{equation}
D_{i,t+1}(x)
=
q_{i,t}^{*}
\left[
\widetilde{\pi}_{i,t}^{B}(x)
-\widetilde{\pi}_{i,t}^{S}(x)
\right].
\label{eq:demand}
\end{equation}
The next-period price is the feasible root closest to \(X_t\):
\begin{equation}
X_{t+1}
=
\underset{x>0}{\arg\min}\;
\left\{|x-X_t|:
\sum_{i=1}^{N}D_{i,t+1}(x)=0
\right\}.
\label{eq:clearing}
\end{equation}
Selecting the closest root avoids discontinuous jumps between multiple
clearing solutions and corresponds to the numerical procedure used in the
simulation.

\section{Information diffusion}
\label{sec:information}

\subsection{Belief updating}

At date \(t\), a signal
\[
I_t=\{x^{*},\sigma^{*},p\}
\]
states that the next-period price is centered at \(x^{*}\) with uncertainty
\(\sigma^{*}\), and the investor assigns probability \(p\) to the signal being
informative. An investor reached by the signal updates the prior in
Eq.~\eqref{eq:prior} to the mixture
\begin{align}
f_{i,t+1}(x\mid X_t,I_t)
={}&
(1-p)
\frac{\exp[-(x-\widehat{x}_i)^2/(2\widehat{\sigma}_i^2)]}
{\sqrt{2\pi}\widehat{\sigma}_i}
\nonumber\\
&+
p
\frac{\exp[-(x-x^{*})^2/(2{\sigma^{*}}^2)]}
{\sqrt{2\pi}\sigma^{*}}.
\label{eq:mixture}
\end{align}
The mixture mean and variance are
\begin{align}
\mu_{i,t+1}^{I}
&=(1-p)\widehat{x}_i+px^{*},
\label{eq:mixturemean}\\
{v}_{i,t+1}^{I}
&=(1-p)\widehat{\sigma}_{i}^{\,2}
+p{\sigma^{*}}^{2}
+p(1-p)(\widehat{x}_i-x^{*})^2.
\label{eq:mixturevariance}
\end{align}
Accordingly, Eq.~\eqref{eq:position} becomes
\begin{equation}
q_{i,t}^{*,I}
=
\frac{\mu_{i,t+1}^{I}-X_t e^{r_f}}
{2\lambda_i {v}_{i,t+1}^{I}}.
\label{eq:updatedposition}
\end{equation}

\subsection{Finite-speed diffusion}

The signal begins at one node and reaches at most \(M\) previously uninformed
neighbors per period. Each agent updates to Eq.~\eqref{eq:mixture} only once
for a given signal. This finite-speed process creates a difference between
investors who have received \(I_t\) and those who still rely on their priors.
It therefore produces gradual price adjustment without requiring herding.

The distinction between the two mechanisms is operational. Information
diffusion changes \(f_{i,t+1}\), and therefore the moments in
Eq.~\eqref{eq:position}; herding changes the action probabilities in
Eq.~\eqref{eq:herding}. Setting \(\gamma=0\) isolates delayed informational
adjustment. Holding beliefs fixed and varying \(\gamma\) isolates social
reinforcement.

\section{Simulation design and mechanisms}
\label{sec:simulation}

\subsection{Calibration}

Table \ref{tab:parameters} reports the baseline parameters inherited from the
simulation study. Heterogeneous priors are drawn across investors, while the
signal is centered above the initial market level. The objective is mechanism
identification rather than structural estimation; consequently, the reported
paths should be interpreted as comparative simulations, not point forecasts.

\begin{table}[htbp]
\centering
\caption{Baseline simulation parameters}
\label{tab:parameters}
\begin{tabularx}{\linewidth}{@{}l l Y@{}}
\toprule
Parameter & Baseline & Interpretation \\
\midrule
\(\Delta\) & \(0.01\) & Half-width of the inaction band \\
\(k\) & \(0.03\) & Sensitivity of action probabilities to price \\
\(\lambda_i\) & \(0.5\) & Risk-aversion coefficient \\
\(d_i\) & \(4\) or \(8\) & Degree under von Neumann or Moore interaction \\
\(\widehat{x}_i\) & \(\mathcal{N}(3000,30^2)\) & Cross-sectional prior means \\
\(\widehat{\sigma}_i\) & \(\mathcal{N}(30,5^2)\), truncated \(>0\) &
Cross-sectional prior standard deviations \\
\(p\) & \(0.6\) & Weight assigned to the information signal \\
\(x^{*}\) & \(3100\) & Signal-implied mean price \\
\(\sigma^{*}\) & \(10\) & Signal uncertainty \\
\(N\) & \(400\) & Number of agents on the \(20\times20\) lattice \\
\bottomrule
\end{tabularx}
\end{table}

\subsection{Herding without new information}

The first experiment holds beliefs fixed and varies
\(\gamma\in\{0.1,0.3,0.5\}\). At low herding intensity, actions remain
spatially dispersed and simulated returns are approximately symmetric. As
\(\gamma\) rises, local majorities persist, trading becomes clustered, and
price changes become more concentrated around zero while extreme observations
occur more often. This combination generates the sharp center and heavy tails
commonly observed in financial returns.

\begin{figure}[hbt]
    \centering
    \begin{subfigure}{0.3\textwidth}
        \includegraphics[width=\textwidth]{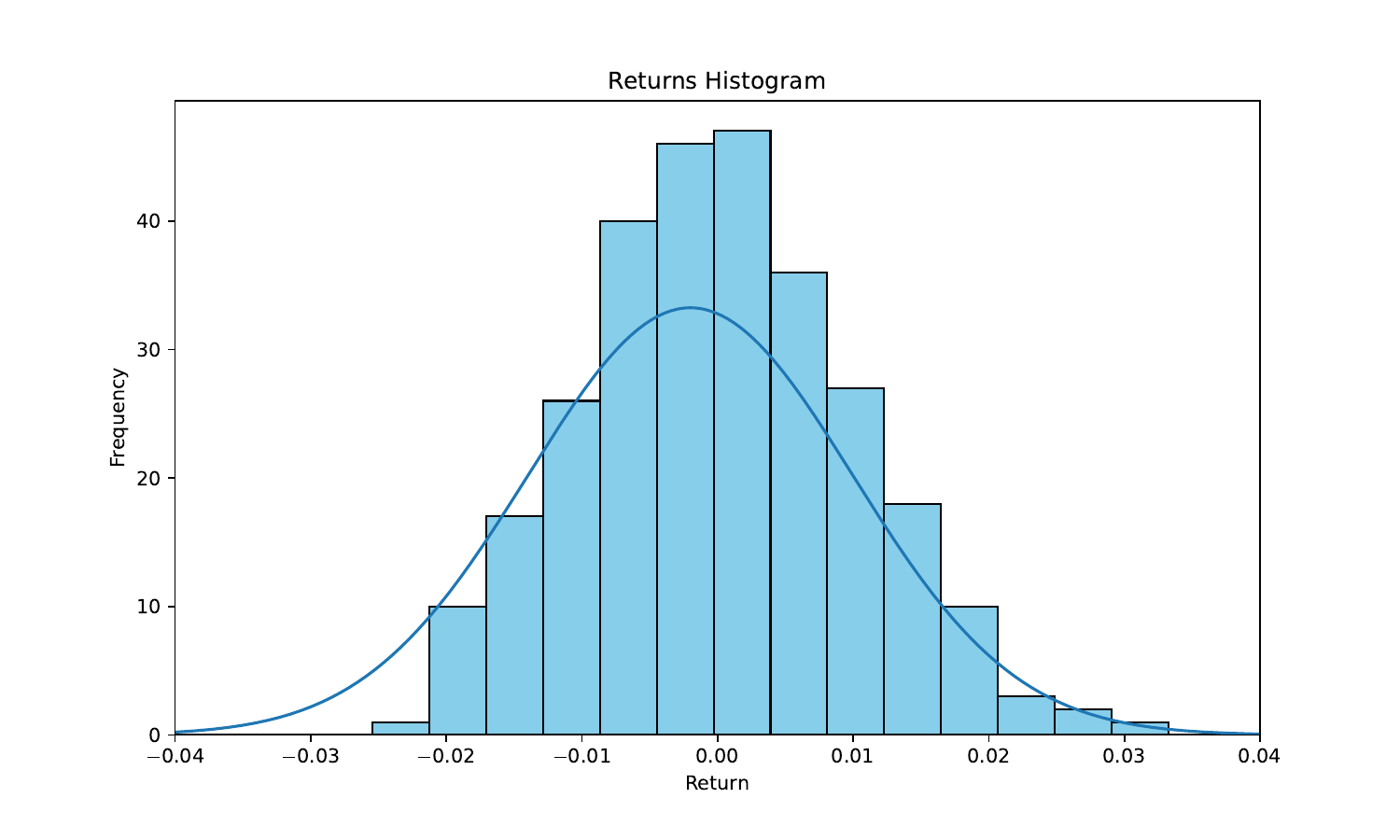}
        \caption{Herding intensity \(\gamma=0.1\)}
        \label{fig:return-gamma-01}
    \end{subfigure}
    \hfill
    \begin{subfigure}{0.3\textwidth}
        \includegraphics[width=\textwidth]{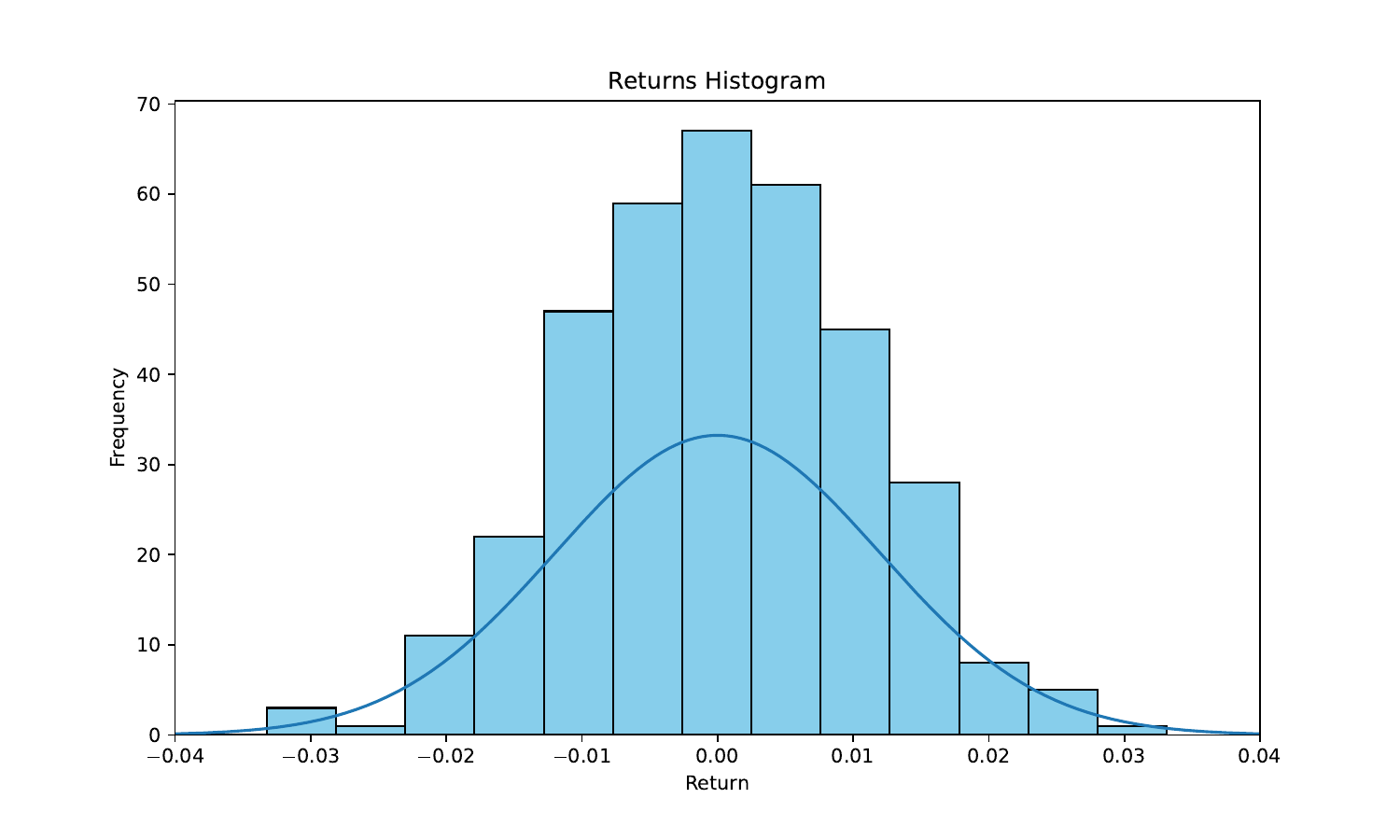}
        \caption{Herding intensity \(\gamma=0.3\)}
        \label{fig:return-gamma-03}
    \end{subfigure}
    \hfill
    \begin{subfigure}{0.3\textwidth}
        \includegraphics[width=\textwidth]{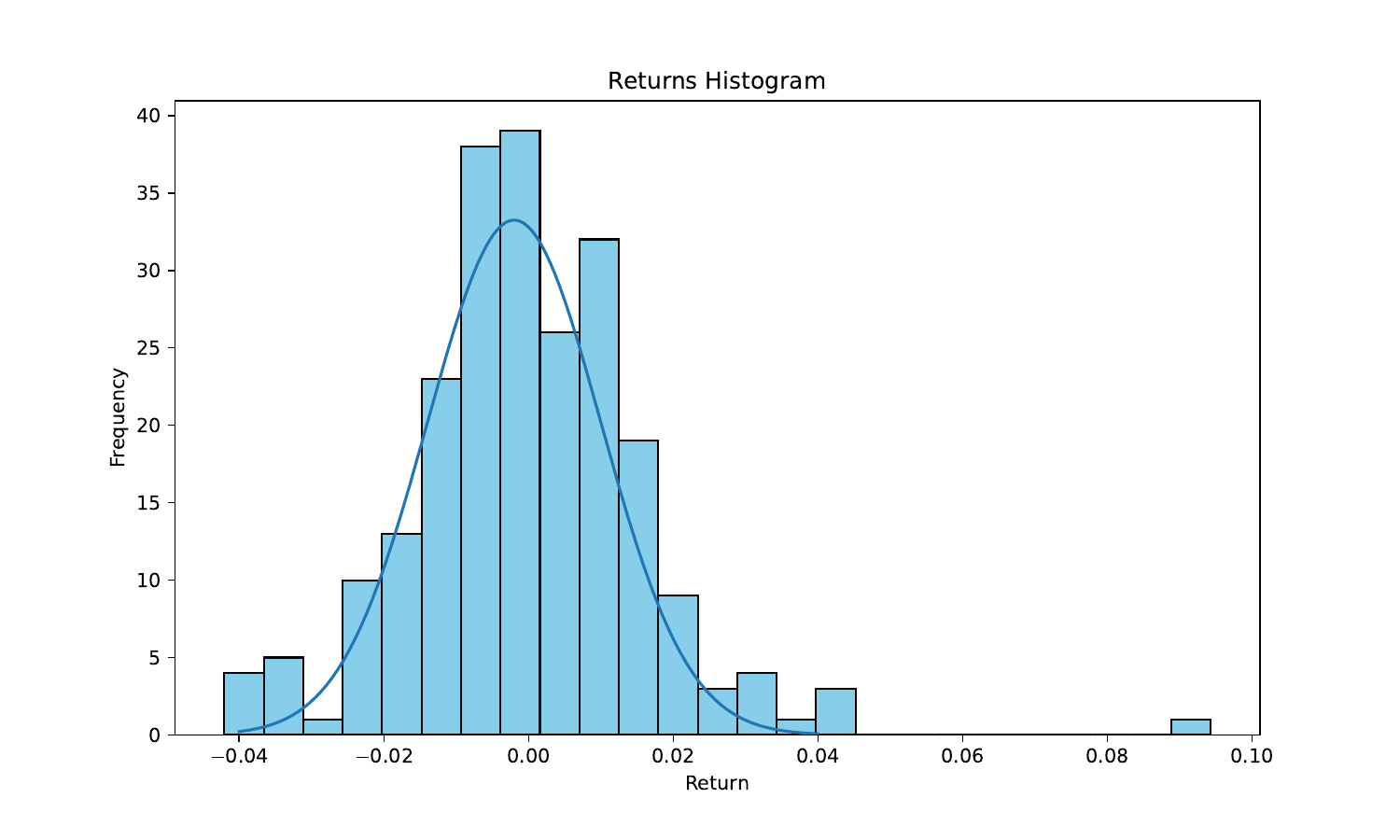}
        \caption{Herding intensity \(\gamma=0.5\)}
        \label{fig:return-gamma-05}
    \end{subfigure}
    \caption{Simulated return distributions over 500 periods under different
    herding intensities.}
    \label{fig:returnhist}
\end{figure}

\begin{figure}[hbt]
    \centering
    \begin{subfigure}{0.3\textwidth}
        \includegraphics[width=\textwidth]{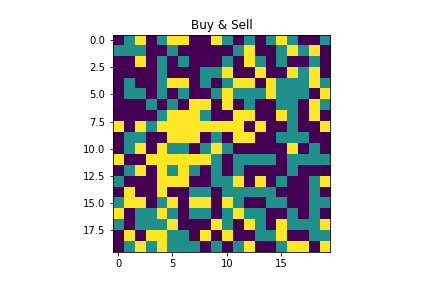}
        \caption{Herding intensity \(\gamma=0.1\)}
        \label{fig:actions-gamma-01}
    \end{subfigure}
    \begin{subfigure}{0.3\textwidth}
        \includegraphics[width=\textwidth]{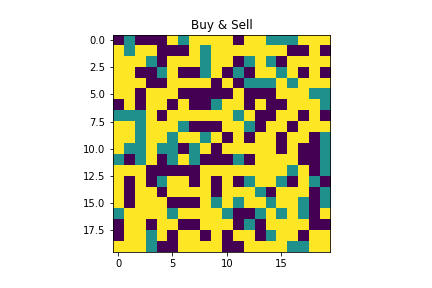}
        \caption{Herding intensity \(\gamma=0.3\)}
        \label{fig:actions-gamma-03}
    \end{subfigure}
    \begin{subfigure}{0.3\textwidth}
        \includegraphics[width=\textwidth]{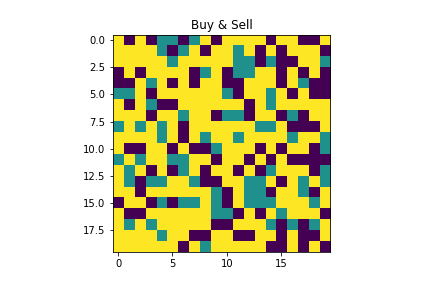}
        \caption{Herding intensity \(\gamma=0.5\)}
        \label{fig:actions-gamma-05}
    \end{subfigure}
    \caption{Cluster trading patterns under different herding intensities.}
    \label{fig:actionmaps}
\end{figure}

\begin{figure}[hbt]
    \centering
    \includegraphics[width=0.5\linewidth]{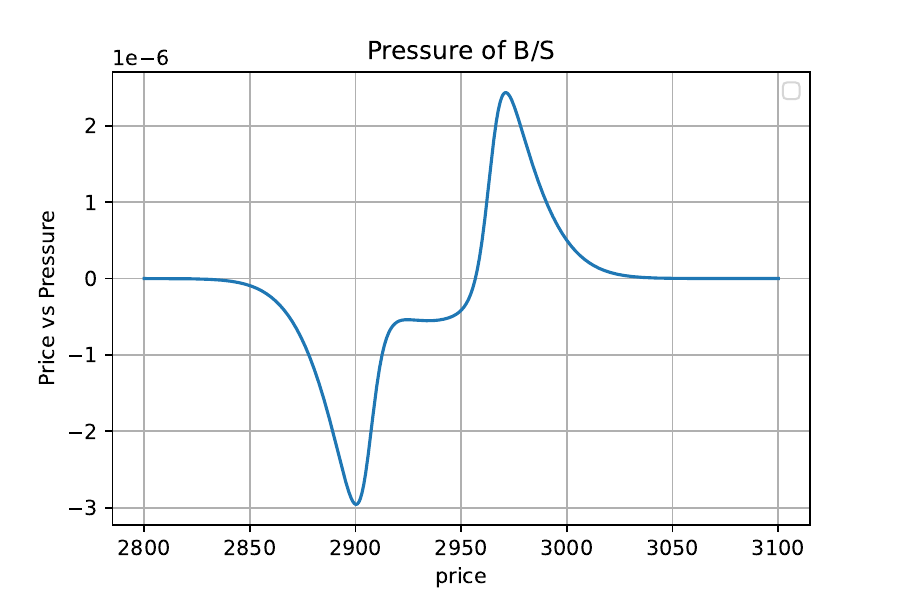}
    \caption{The equilibrium price is the zero of the buy--sell pressure
    function closest to the previous price.}
    \label{fig:clearing}
\end{figure}

\subsection{Information diffusion without herding}

The second experiment sets \(\gamma=0\) and activates the signal. Prices move
toward \(x^{*}\) as informed beliefs propagate across the network. Because
information reaches only part of the market in each period, the adjustment is
gradual and produces short-horizon continuation. Network degree governs the
speed of convergence. In repeated \(20\times20\) simulations, the reported mean
time to reach the signal mean is 15.41 periods for the degree-four von Neumann
lattice and 9.22 periods for the degree-eight Moore lattice.

Overshooting can occur even in the absence of herding when the temporary
imbalance between informed and uninformed demand pushes the clearing price past
the signal mean. As diffusion completes and beliefs become less heterogeneous,
the price returns toward the signal-implied region. This experiment shows that
momentum and reversal can arise from the information process alone; herding is
an amplifier rather than a necessary condition.

\begin{figure}[hbt]
    \centering
    \begin{subfigure}{0.3\textwidth}
        \includegraphics[width=\textwidth]{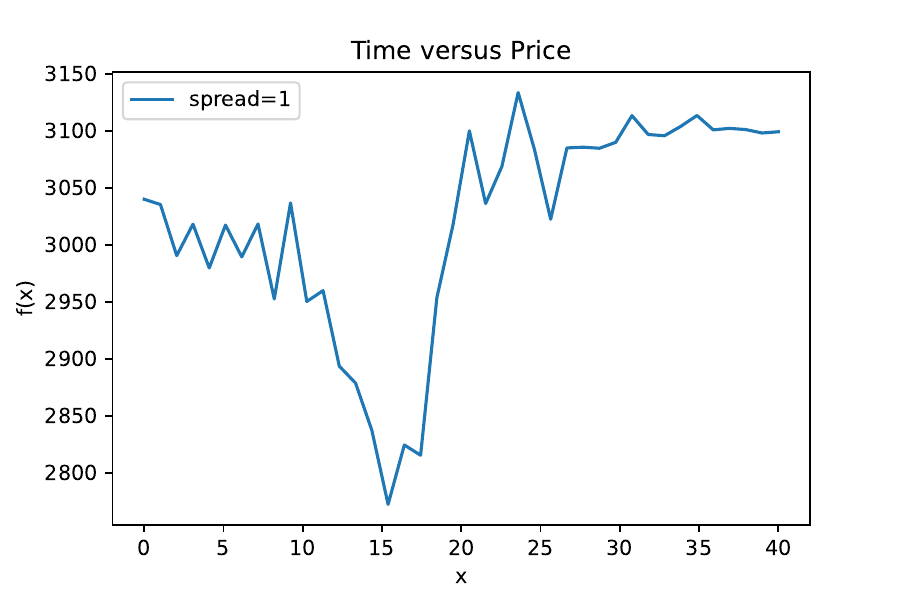}
        \caption{Von Neumann}
        \label{fig:information-von-neumann}
    \end{subfigure}
    \hfill
    \begin{subfigure}{0.3\textwidth}
        \includegraphics[width=\textwidth]{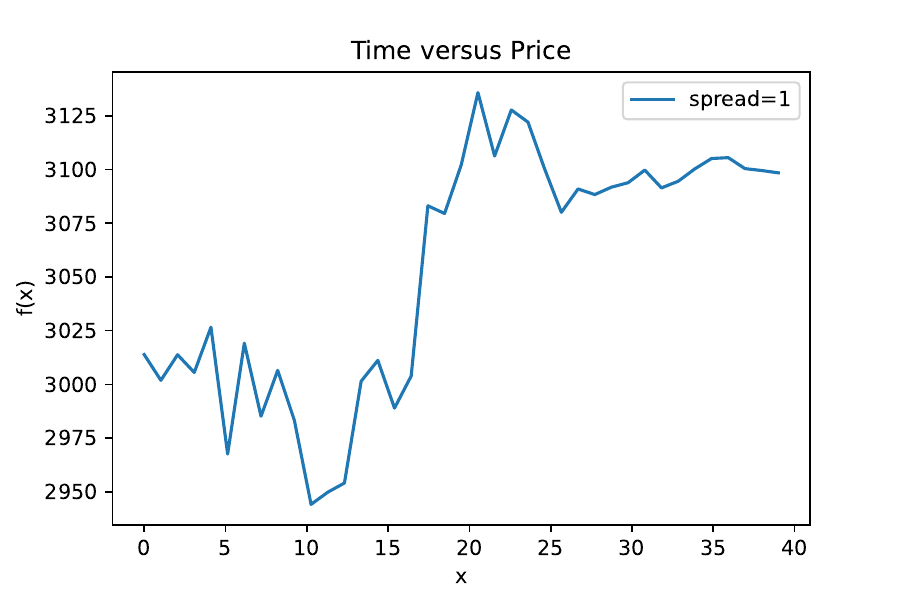}
        \caption{Komogelov}
        \label{fig:information-komogelov}
    \end{subfigure}
    \hfill
    \begin{subfigure}{0.3\textwidth}
        \includegraphics[width=\textwidth]{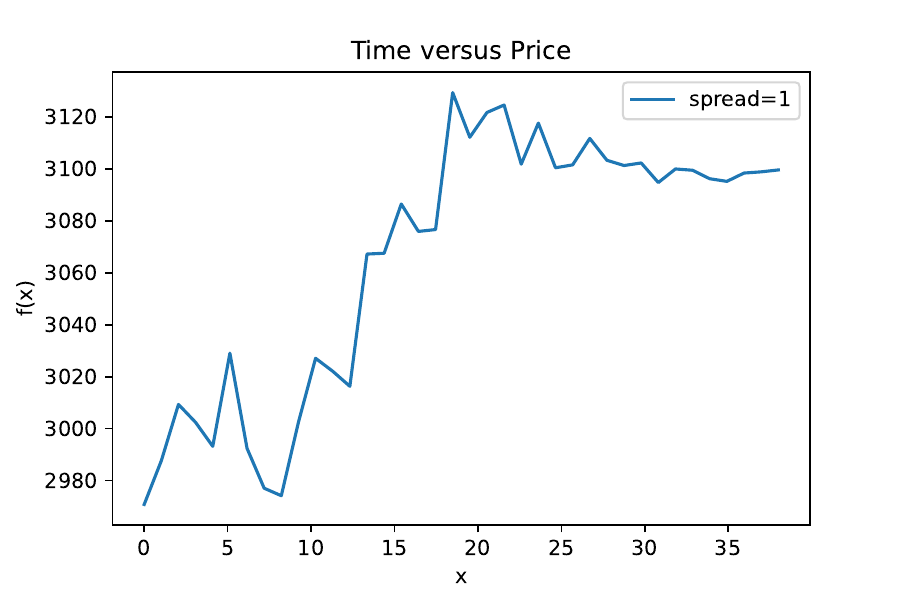}
        \caption{Moore}
        \label{fig:information-moore}
    \end{subfigure}
    \caption{Possible price paths driven only by information changes under
    different regular networks.}
    \label{fig:informationpaths}
\end{figure}

\subsection{Joint dynamics: momentum and reversal}

When information diffusion and herding operate jointly, early recipients of a
positive signal buy first. Their actions raise both the price and the local buy
shares observed by neighbors. Newly informed investors respond to the signal,
while uninformed investors may imitate the observed trades. This positive
feedback strengthens return continuation relative to the information-only
case. The mechanism is consistent with gradual-information models
\citep{hong1999}, but it makes the social channel explicit.

The same feedback also creates the conditions for reversal. Once most agents
have adopted the dominant action, further reinforcement is limited. If the
market price has moved beyond the signal-implied value, updated expected
returns in Eq.~\eqref{eq:updatedposition} turn negative for marginal buyers.
The weakening or reversal of the local majority then produces coordinated
selling. Thus, reversal follows from the combination of overshooting, herding
saturation, and subsequent belief correction.

\begin{figure}[hbt]
    \centering
    \begin{subfigure}{0.4\textwidth}
        \includegraphics[width=\textwidth]{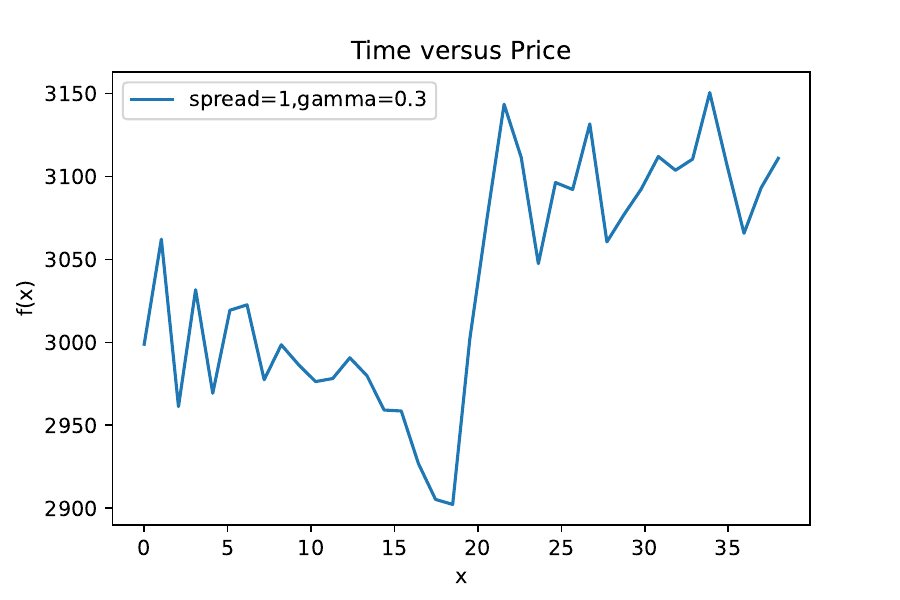}
        \caption{Von Neumann}
        \label{fig:joint-von-neumann}
    \end{subfigure}
    \begin{subfigure}{0.4\textwidth}
        \includegraphics[width=\textwidth]{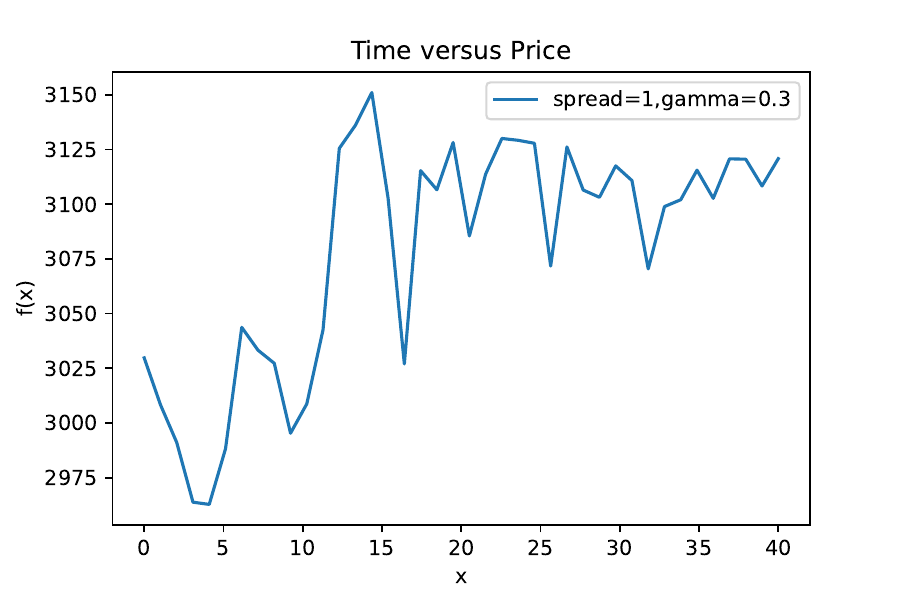}
        \caption{Moore}
        \label{fig:joint-moore}
    \end{subfigure}
    \caption{Information diffusion with herding produces stronger
    fluctuations near the signal-implied price than information alone.}
    \label{fig:jointpaths}
\end{figure}

\begin{figure}[hbt]
    \centering
    \includegraphics[width=0.4\linewidth]{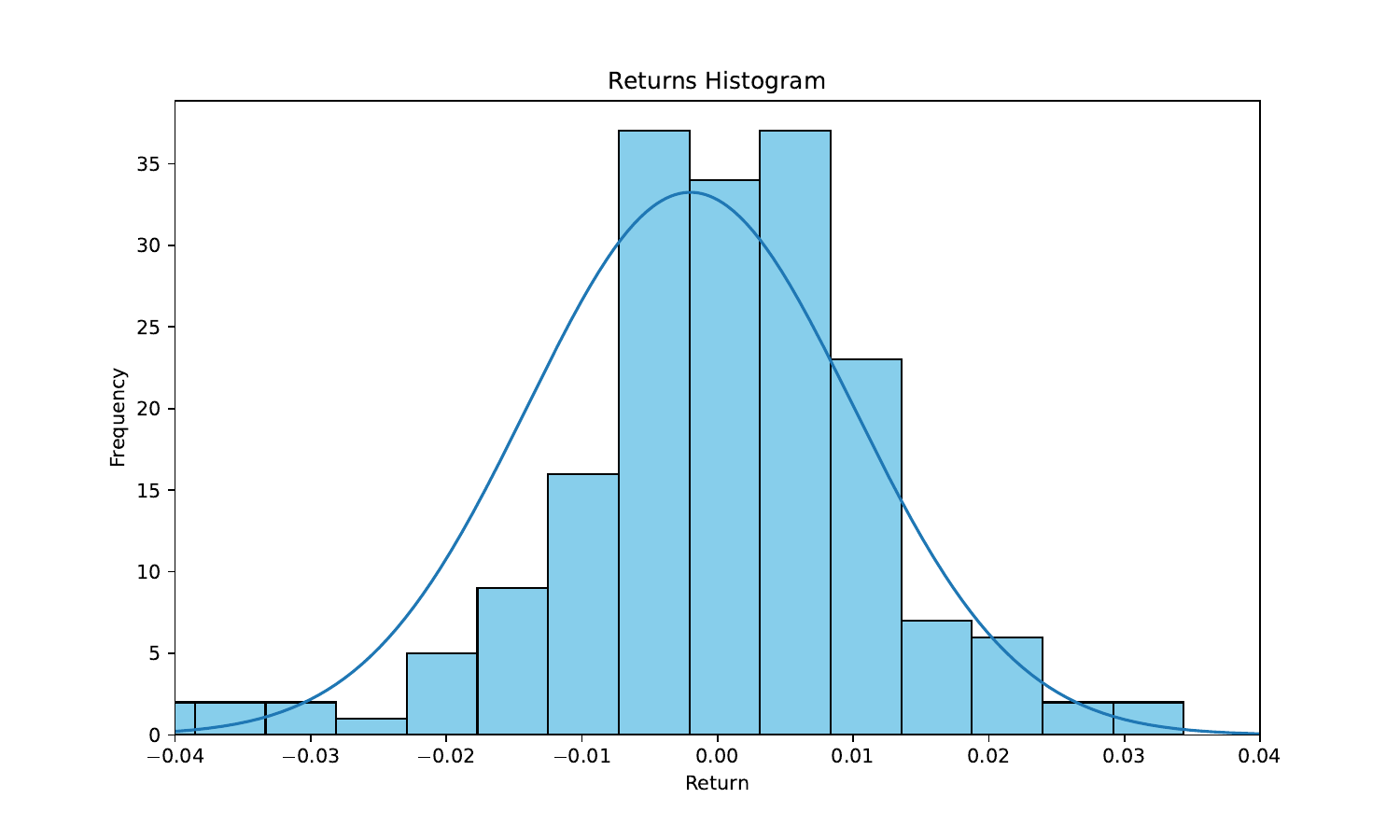}
    \caption{Return distribution under joint information diffusion and herding.}
    \label{fig:joint-return-distribution}
\end{figure}

\section{Empirical application}
\label{sec:empirical}


\subsection{Conventional herding measures}

The empirical application uses a rolling sample from China's A-share market
over 2015--2024, with the SSE 50 universe used for the market-level comparison.
Let \(R_{i,t}\) be stock \(i\)'s return and \(R_{m,t}\) the cross-sectional
market return. The cross-sectional absolute deviation is
\begin{equation}
\operatorname{CSAD}_{t}
=
\frac{1}{n_t}
\sum_{i=1}^{n_t}
\left|R_{i,t}-R_{m,t}\right|.
\label{eq:csad}
\end{equation}
Following \citet{chang2000}, aggregate herding is tested with
\begin{equation}
\operatorname{CSAD}_{t}
=
\alpha
+\beta_1|R_{m,t}|
+\beta_2R_{m,t}^{2}
+\varepsilon_t.
\label{eq:cck}
\end{equation}
A significantly negative \(\beta_2\) indicates that dispersion increases less
than proportionally during large market movements, consistent with correlated
trading.

For trade-based analysis, let \(B_{i,t}\) and \(S_{i,t}\) denote classified buy
and sell counts, \(N_{i,t}=B_{i,t}+S_{i,t}\), and
\(p_t=\sum_i B_{i,t}/\sum_iN_{i,t}\). The LSV measure is
\begin{equation}
\operatorname{LSV}_{i,t}
=
\left|
\frac{B_{i,t}}{N_{i,t}}-p_t
\right|
-\operatorname{AF}_{i,t},
\label{eq:lsv}
\end{equation}
where
\begin{equation}
\operatorname{AF}_{i,t}
=
\mathbb{E}
\left[
\left|
\frac{\widetilde{B}_{i,t}}{N_{i,t}}-p_t
\right|
\right],
\qquad
\widetilde{B}_{i,t}\sim
\operatorname{Binomial}(N_{i,t},p_t),
\label{eq:adjustment}
\end{equation}
is the finite-sample adjustment under independent trading.

\subsection{A transformed-tail indicator}
\label{sec:tailindicator}

The simulations suggest that social reinforcement raises the mass in both
return tails, while directional information can introduce skewness. To reduce
the influence of skewness before measuring tail thickness, returns in each
30-trading-day window are mapped to a Johnson \(S_U\) variable
\begin{equation}
z_t
=
a+b\,\operatorname{asinh}
\left(
\frac{r_t-c}{d}
\right),
\qquad b,d>0,
\label{eq:johnson}
\end{equation}
where \(a,b,c,d\) are fitted within the window. After standardization, the
tail-based herding statistic at threshold \(u>0\) is
\begin{equation}
\operatorname{HerdTail}_{t}(u)
=
\widehat{\Pr}(z<-u)
+\widehat{\Pr}(z>u)
-2[1-\Phi(u)],
\label{eq:tailindicator}
\end{equation}
where \(\Phi\) is the standard normal distribution function. Positive values
indicate more two-sided tail mass than the Gaussian benchmark.

The rolling indicator broadly tracks the aggregate LSV series in the supplied
analysis. Both measures rise around the 2015 market disruption and the
COVID-19 episode, which is consistent with stronger correlated trading in
stress periods. This evidence should be interpreted as validation of
co-movement, not as causal identification: volatility, common news, liquidity
constraints, and price limits can also increase tail mass.

\begin{figure}[hbt]
    \centering
    \includegraphics[width=0.9\linewidth]{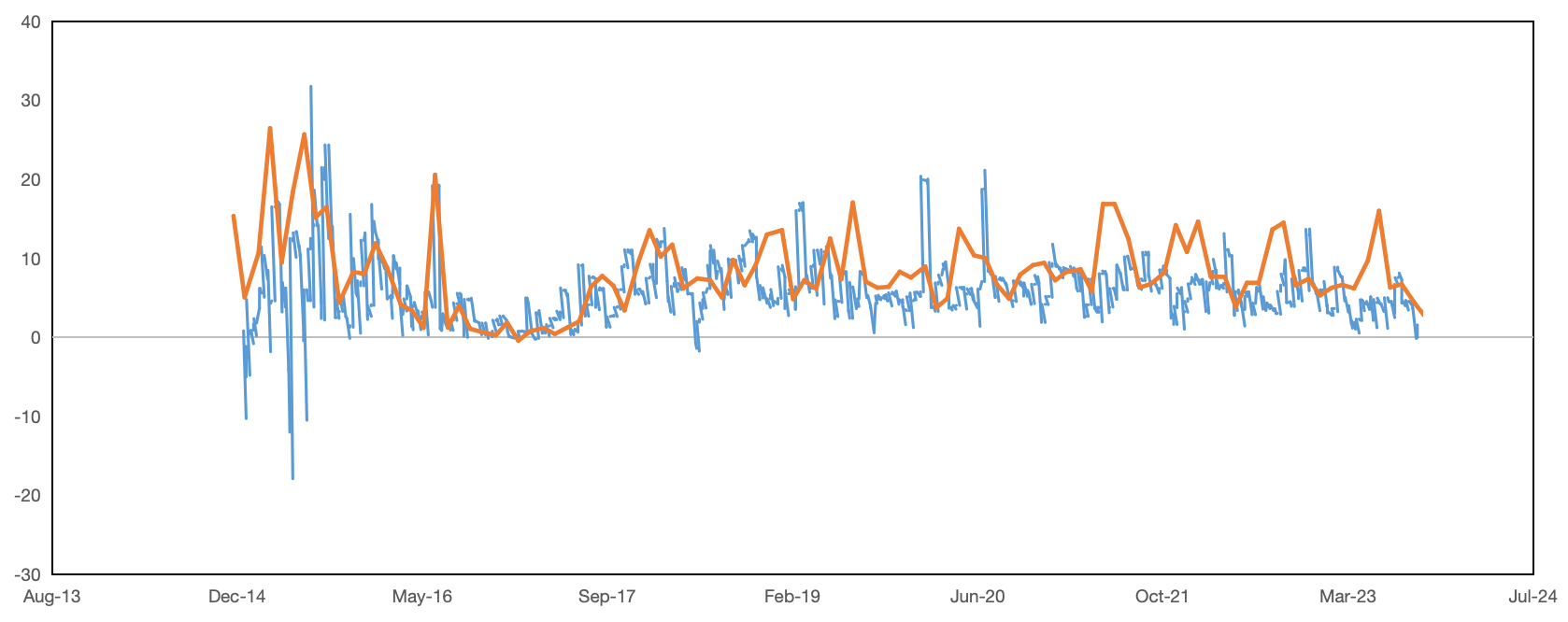}
    \caption{Comparison between the LSV measure and the return-distribution
    measure.}
    \label{fig:lsvcomparison}
\end{figure}

\subsection{Trading-rule illustration}

The original simulation study also considers a directional rule that combines
the herding indicator with the recent price trend. In an uptrend, the rule
maintains the position while herding is strong and exits when the indicator
weakens; in a downtrend, it enters only after herding begins to dissipate. The
reported 2019--2024 path outperforms the market benchmark. Because the current
analysis does not yet report transaction costs, turnover, parameter-selection
protocols, or a fully held-out test period, the result is presented as an
illustration rather than evidence of an implementable anomaly.

\begin{figure} [bthp]
    \centering
    \includegraphics[width=0.5\linewidth]{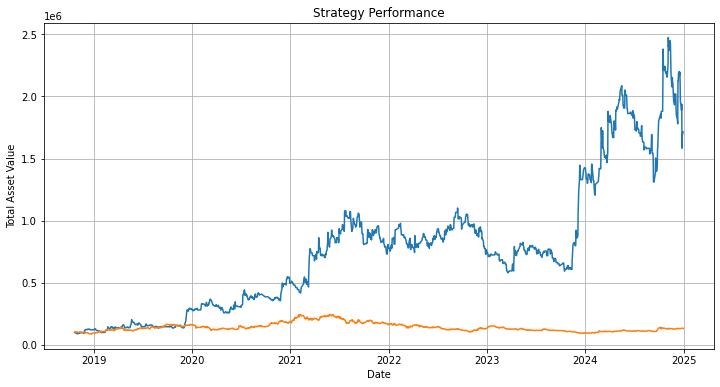}
    \caption{The strategy return exceeds the market return over 2019--2024.}
    \label{fig:out}
\end{figure}

\section{Network robustness}
\label{sec:robustness}

Regular lattices hold degree constant and make the diffusion mechanism
transparent, but actual investor networks contain random links, hubs, and short
paths. Two alternatives are therefore considered. In an
Erd\H{o}s--R\'enyi network, each potential link is formed randomly
\citep{erdos1959}. In a
Watts--Strogatz network, a locally connected ring is rewired with probability
\(\rho\), which preserves clustering while shortening average path length
\citep{watts1998}.

Repeating the simulation on both networks preserves the central qualitative
results: herding produces clustered decisions and excess return tails, delayed
information generates gradual adjustment, and their combination can generate
overshooting followed by reversal. Topology affects the speed and concentration
of the response, but not the sign of the mechanism.

\begin{figure}[hbt]
    \centering
    \begin{subfigure}{0.25\textwidth}
        \includegraphics[width=\textwidth]{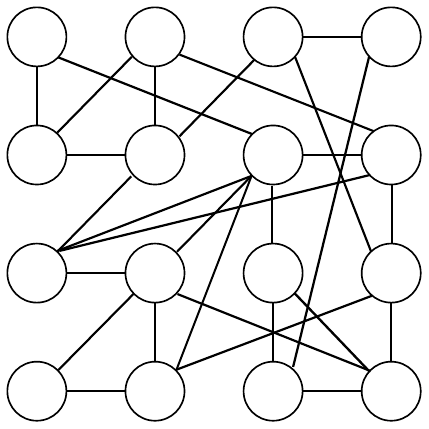}
        \caption{Erd\H{o}s--R\'enyi network}
        \label{fig:er-network}
    \end{subfigure}
    \begin{subfigure}{0.25\textwidth}
        \includegraphics[width=\textwidth]{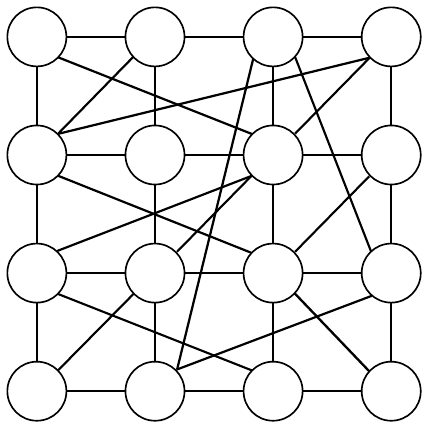}
        \caption{Small-world network}
        \label{fig:small-world-network}
    \end{subfigure}
    \caption{Two random networks with \(4\times4\) nodes.}
    \label{fig:random-networks}
\end{figure}

\begin{figure}[hbt]
    \centering
    \begin{subfigure}{0.4\textwidth}
        \includegraphics[width=\textwidth]{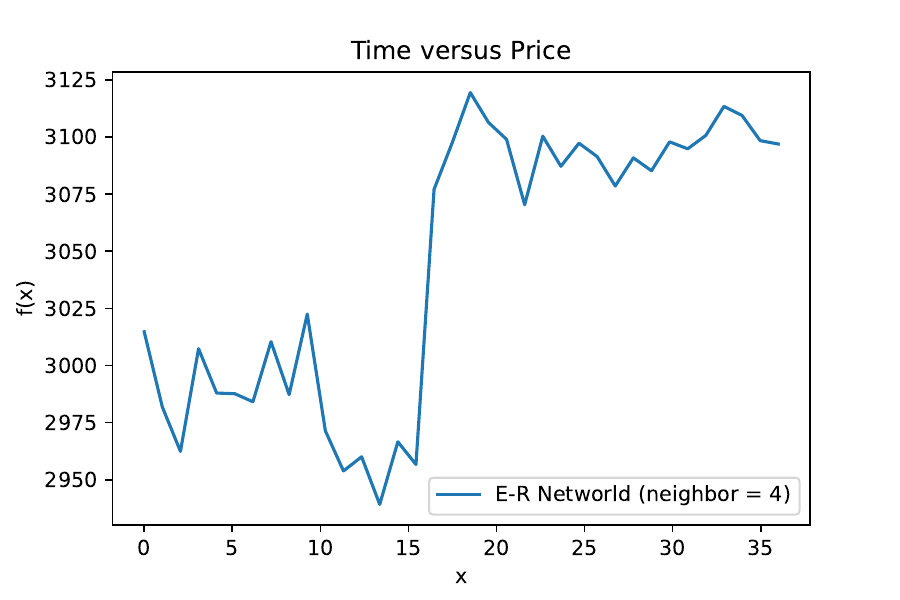}
        \caption{A possible price path on the E-R network}
        \label{fig:er-price-path}
    \end{subfigure}
    \begin{subfigure}{0.4\textwidth}
        \includegraphics[width=\textwidth]{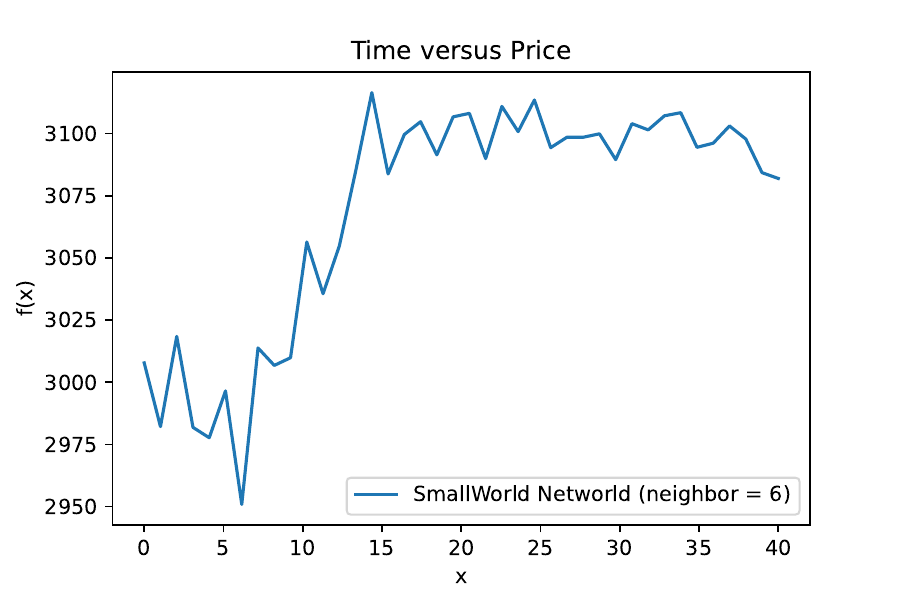}
        \caption{A possible price path on the small-world network}
        \label{fig:sw-price-path}
    \end{subfigure}
    \begin{subfigure}{0.4\textwidth}
        \includegraphics[width=\textwidth]{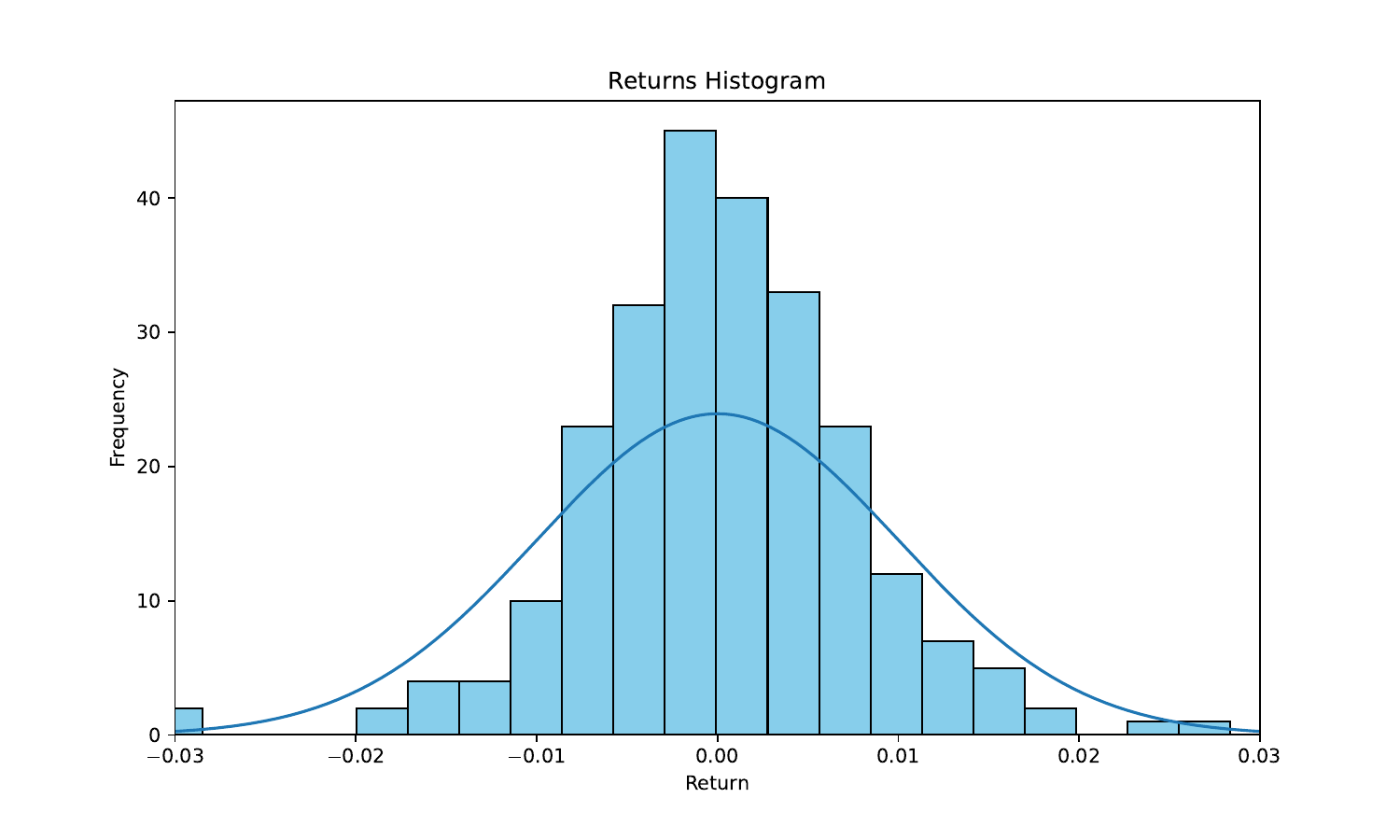}
        \caption{Return distribution on the E-R network}
        \label{fig:er-return-distribution}
    \end{subfigure}
    \begin{subfigure}{0.4\textwidth}
        \includegraphics[width=\textwidth]{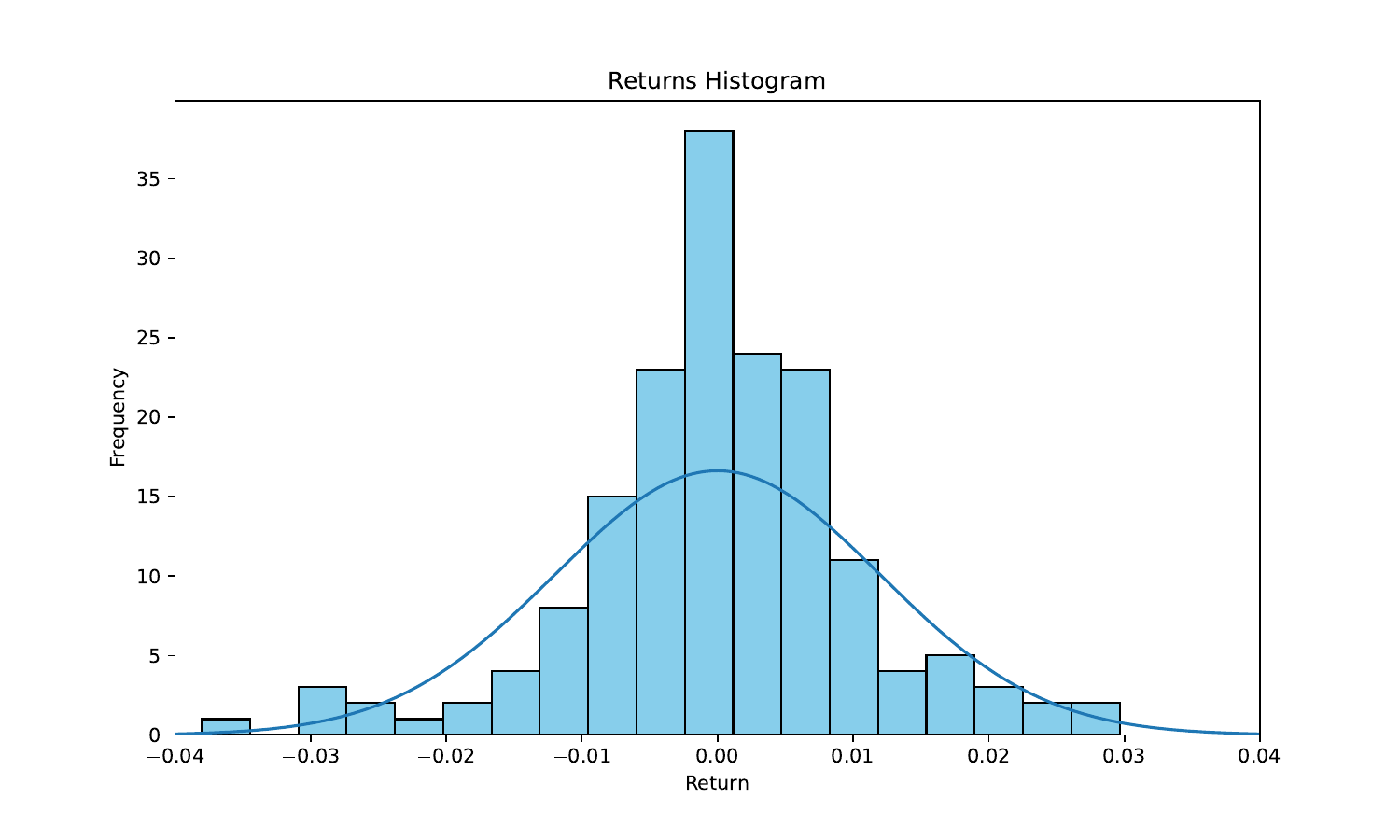}
        \caption{Return distribution on the small-world network}
        \label{fig:sw-return-distribution}
    \end{subfigure}
    \caption{Artificial financial market simulations on random networks.}
    \label{fig:networkrobustness}
\end{figure}

\section{Discussion}
\label{sec:discussion}

The model clarifies three conceptually distinct sources of price persistence.
First, slow information arrival generates rational underreaction. Second,
imitation transmits early trades beyond the set of informed investors. Third,
market clearing translates correlated actions into prices, making observed
returns an additional coordination signal. Reversal occurs when these forces
weaken and positions become inconsistent with updated expected payoffs.

The results also explain why empirical herding measures should be interpreted
carefully. A low CSAD or a high LSV value is consistent with correlated
decisions, but it does not identify the underlying motive. The network model
shows that a common signal and imitation may generate similar aggregate
patterns. The transformed-tail statistic has the same limitation; its value is
that it can be computed in a short rolling window and compared with
trade-based measures.

Several limitations remain. The Gaussian prior and three-action rule simplify
belief formation, position constraints, wealth dynamics, and order-book
microstructure. The network is exogenous, whereas real trading relationships
co-evolve with performance and information. The empirical application needs a
fully documented data-construction pipeline, formal uncertainty estimates for
the tail indicator, and out-of-sample evaluation of the trading rule. Future
work could estimate network exposure from account-level or social-interaction
data, endogenize link formation, and compare the model against alternatives
using simulated-method-of-moments or likelihood-free inference.

\section{Conclusion}
\label{sec:conclusion}

This paper develops a network-based artificial financial market that separates
belief updating from social imitation. Delayed information diffusion explains
gradual price adjustment; local herding amplifies the associated trend and
produces clustered trading and heavy-tailed returns. When the local majority
saturates and beliefs catch up with the signal, the same mechanism can produce
overshooting and reversal.

The empirical application complements conventional CSAD and LSV measures with
a rolling, transformed-tail indicator. Their similar movements in the A-share
sample support the model's central implication that concentrated trading is
associated with non-Gaussian return behavior. The evidence is descriptive and
does not rule out common-information or liquidity channels, but it demonstrates
how network interaction can connect micro-level imitation to market-level
momentum and reversal.

\section*{Declaration of generative AI and AI-assisted technologies in the
manuscript preparation process}

OpenAI Codex was used to translate the source manuscript, reorganize the
article, edit the English, and prepare the LaTeX source. The author reviewed and
revised the generated material and accepts full responsibility for the content
of the article.

\bibliographystyle{elsarticle-harv}
\bibliography{references_elsevier}

\end{document}